\documentclass[a4paper]{jpconf}
\usepackage{graphicx}
\usepackage{citesort}
\newcommand{\micron}{\mbox{$\mu$}m}
\newcommand{\lbol}{$L_{\rm bol}$}
\def\arcsec{\hbox{$^{\prime\prime}$}}
\def\degr{\hbox{$^\circ$}}
\def\araa{ARA\&A}%
\def\apj{ApJ}%
\def\apjl{ApJ}%
\def\apjs{ApJS}%
\def\aap{A\&A}%
\def\mnras{MNRAS}%
\def\mathstacksym#1#2#3#4#5{\def#1{\mathrel{\hbox to 0pt{\lower#5\hbox{#3}\hss} \raise #4\hbox{#2}}}}

\mathstacksym\gta{$>$}{$\sim$}{1.5pt}{3.5pt} 
\mathstacksym\lta{$<$}{$\sim$}{1.5pt}{3.5pt} 

\begin{document}
\title{Young massive stars and their environment in the mid-infrared at high angular resolution}

\author{W.J. de Wit$^1$, M.G. Hoare$^1$, R.D. Oudmaijer$^1$, T. Fujiyoshi$^2$}
\address{$^1$ School of Physics \& Astronomy, University of Leeds, LS2 9JT, UK}
\address{$^2$ Subaru Telescope, NAOJ, 650 North A'ohoku Place, Hilo, HI 96720, USA}

\ead{w.j.m.dewit@leeds.ac.uk}

\begin{abstract}
We present interferometric and single-dish mid-infrared observations
of a sample of massive young stellar objects (BN-type objects), using
VLTI-MIDI (10\,\micron) and Subaru-COMICS (24.5\,\micron). We discuss the
regions S140, Mon R2, M8E-IR, and W33A. The observations probe the
inner regions of the dusty envelope at scales of 50 milli arcsecond
and 0.6\arcsec ($\sim$100-1000\,AU), respectively. Simultaneous model fits to spectral
energy distributions and spatial data are achieved using
self-consistent spherical envelope modelling. We conclude that those
MYSO envelopes that are best described by a spherical geometry, the
commensurate density distribution is a powerlaw with index $-1.0$. Such 
a powerlaw is predicted if the envelope is supported by turbulence 
on the 100-1000\,AU scales probed with MIDI and COMICS, but the r\^{o}le
of rotation at these spatial scales need testing.

\end{abstract}

\section{Introduction}
Massive stars ($>8 M_{\odot}$) play a major role in the dynamical, thermal and
chemical evolution of galaxies. The question of how massive stars form is
therefore central for a proper understanding of these processes. Currently
there is no consensus regarding the answer to this question, and two diverging
viewpoints regarding the initial formation stages are generally referred
to. On the one hand, a high-mass star is formed from the monolithic
gravitational collapse of a massive molecular condensation (core). The
principal stellar product is a single or binary massive star that is formed
from gravitationally bound core material \cite{2003ApJ...585..850M} and accreted
onto the central star through a disk. The competing viewpoint is due to
\cite{1997MNRAS.285..201B} in which a massive core fragments into a collection
of equal mass stellar seeds that compete via Bondi-Hoyle accretion for
inflowing material. The seed that ends up being a massive star happened to be
located in a fortuitous spatial position (near the centre of the potential
well) where most of the inflowing material gathered. The amount of
observational evidence \cite{2007ApJ...654..304K,2007A&A...476.1243M,2007A&A...462L..17A,2006MNRAS.367..737B} and theoretical
considerations \cite{2004MNRAS.349..678E,2006ApJ...641L..45K} that favour the monolithic collapse scenario is growing. 
Recent developments in massive star formation are reviewed in \cite{2007ARA&A..45..481Z,2007ARA&A..45..565M}.


Whichever scenario dictates massive star formation (SF), they both start out with
a massive and cold molecular core that is gravitationally unstable. Hidden from
direct view by the surrounding dusty envelope, a massive star forms at the
centre of the cold core and its radiation is reprocessed to longer wavelengths
before it escapes the structure. Observations in the infrared (IR) and (sub)mm of star
forming regions in the solar neighbourhood have identified cold objects with
bolometric luminosities equalling that of a (single) massive star, i.e. a few
times $\rm 10^{4}\,L_{\odot}$. This luminosity is such that the central object has the ability
to ionise its surroundings if due to a single main sequence star, yet only
little (if any) recombination emission is observed from these sources. These
objects are identified with the massive young stellar object (MYSO) evolutionary phase of
SF in which the formation of a central massive star is nearing its completion
but the central object still experiences mass accretion. The latter is
evidenced by the ubiquitous bipolar outflow signatures associated with these star
forming regions \cite{2005ccsf.conf..105B}. The star itself, although
unmistakenly present, is not observable as the surrounding dusty 
envelope is highly optically thick ($A_{v}$ up to 50 to 100).  However crucial
clues to the formation process can be conveyed by the radial structure of the
dust envelope. It is determined by the forces that operate during the onset
and the subsequent evolution of the initial molecular core. For example, the
radial density profile is predicted to have a powerlaw  with a value for the
power index that depends on the dominant physics. The envelope emission is a
direct consequence of the dust radial density distribution and the exact power
index can be extracted from the observables by the use of radiative transfer 
models.  Spherical envelope geometries can be assumed for the dust emission
dominating at wavelengths larger than $30\mu$m, where the SEDs of MYSOs are 
observed to be remarkably similar, arguing for little deviation from spherical 
symmetry
\cite{1986A&A...167..315C,1990ApJ...354..247C,1991MNRAS.251..584H,1991A&A...252..801G,1994ApJ...427..889W,1998ApJ...500..280F,2000A&A...357..637H,2000ApJ...537..283V,2002ApJS..143..469M,2002ApJ...566..945B}. 
Emission of MYSOs at shorter wavelengths poses more serious interpretation problems.
 
Here we present an investigation into the structure of dusty MYSO envelopes
at intermediate and high angular resolution in the mid-IR. We discuss
Subaru-COMICS observations at 24.5\,\micron~with an angular resolution of 0.6\arcsec~of a sample of 14 MYSOs, and VLTI-MIDI observation at 10\,\micron~of
1 MYSO with a resolution of 50 milli arcseconds.

\section{Observations and method of analysis} 
Observations at 24.5\,\micron~were taken with the mid-infrared imaging
spectrometer COMICS \cite{2000SPIE.4008.1144K} mounted on the 8.2m Subaru
telescope on Mauna Kea. The system attains a diffraction limited angular
resolution of 0.6\arcsec~and delivers over-sampled images with $\rm
0.13\times0.13\,arcsec^2$ pixels. A total of 14 well-known young massive stars
were observed in a snapshot mode between 2003 and 2006. More details on
observations and data reduction can be found in de Wit et al. (2008 subm.). The
spatially resolved Subaru observations were used to construct an azimuthally
averaged intensity profile as function of distance from the centre of the
object.

The highest resolution observation presented here (50\,mas) were taken with
the MIDI instrument on the VLT interferometer. MIDI delivers
spectro-interferometric information over the wavelength interval
8-13\,\micron. This set consists of 1 baseline on the MYSO W33A. The observations
were taken using the UT2-UT3 telescope configuration for a projected baseline
of $\rm 45.5\,m$ and position angle of $47\degr$ east of north; this is
perpendicular to the star's outflow direction. More details regarding
observations and reduction can be found in \cite{2007ApJ...671L.169D}. The
spatially resolved MIDI observations deliver the size scale of W33A as
function of wavelength.
 
The analysis of the Subaru and MIDI data were quite analogous. Spatial
information is combined with the SED, and we
attempt to simultaneously model these observational data with spherical dust
radiative transfer models. The spherical models were calculated with DUSTY,
a code that solves the 1-D scaled radiative transfer problem
\cite{1997MNRAS.287..799I}. SEDs were built from literature data. The Subaru
sample was compared to a grid of 120\,000 DUSTY models, whereas the best model
for W33A was found iteratively.

We present the results regarding MYSO envelope emission going from the outside
(Subaru) in (MIDI). First, we discuss the type of morphology observed in the
resolved Subaru images.

\section{The 24.5$\mu$m~morphology of MYSOs} 
The COMICS images show that the
main emission component at 24.5\,\micron~is found in discrete single sources
down to the angular resolution limit. These sources are invariably identified
with the known MYSOs of each particular region. The majority of sources in our
sample shows to first order symmetric 24.5\,\micron~emission. In three cases
the discrete sources are resolved in multiple condensations located within a
resolved envelope (AFGL\,961, AFGL\,4029, W3). Extended diffuse emission is
observed in many regions and is associated with either UCHII regions or with
shock excited material. The systematic difference in morphology between UCHII
and MYSOs demonstrates what is likely to be a difference in evolutionary phase: the
break-out of ionised gas of the more evolved UCHII has lifted the dusty
core remnant. The following subsections describe two examples of
discrete and extended 24.5\,\micron~emission as observed with COMICS.

\begin{figure}
  \includegraphics[width=7.9cm,height=8.5cm]{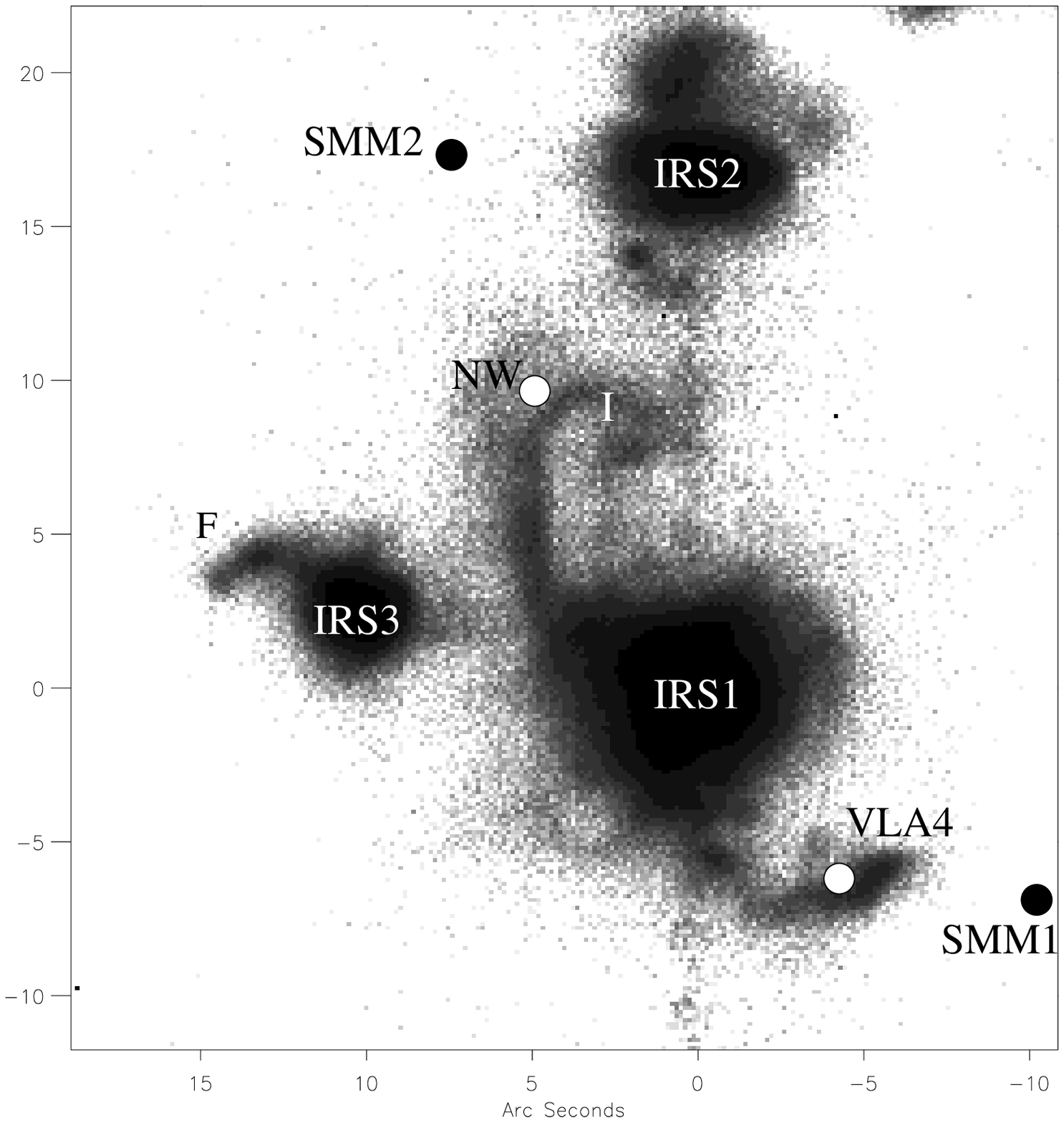}
  \includegraphics[width=7.9cm,height=7.5cm]{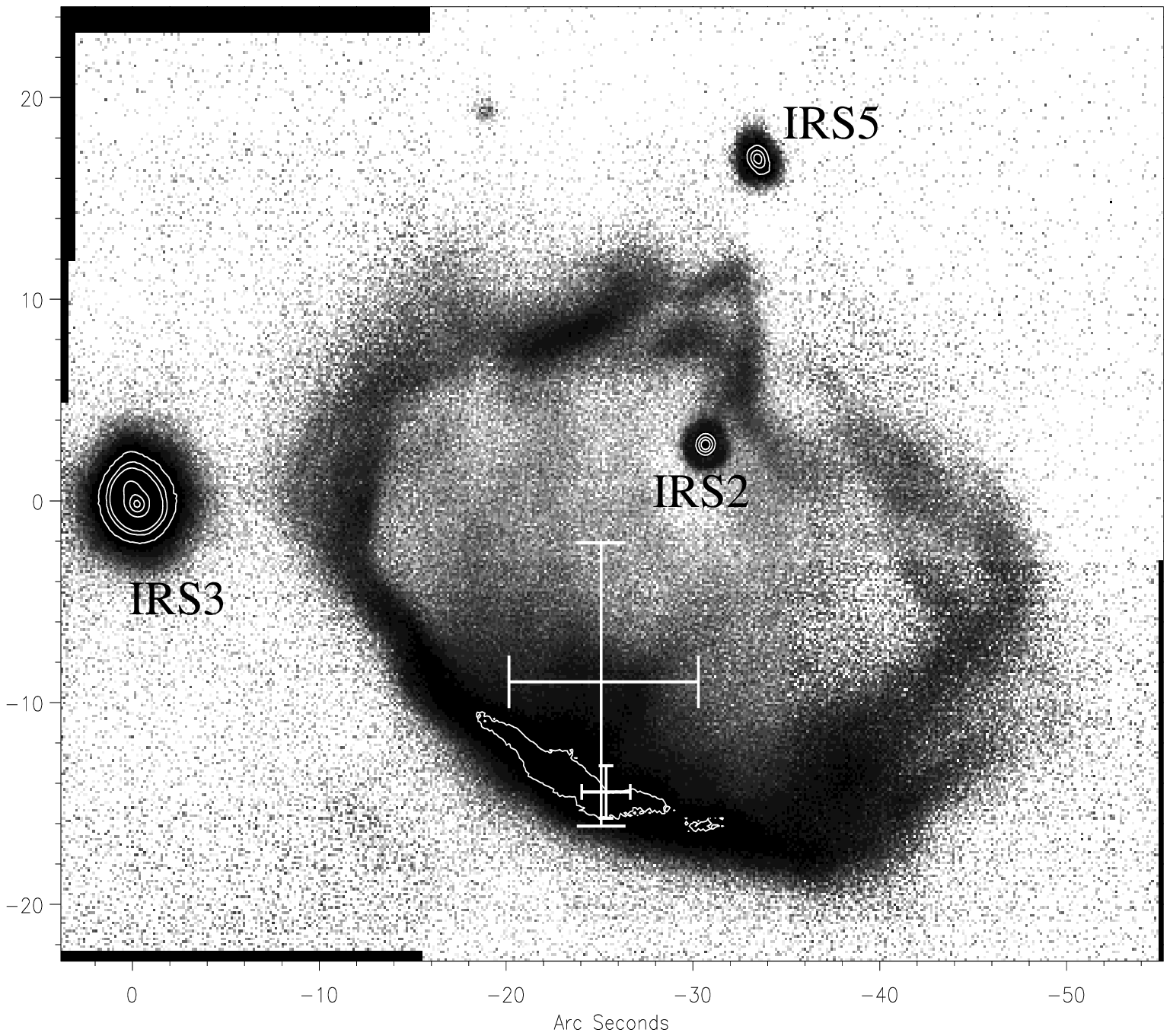}
\caption[]{COMICS images at 0.6\arcsec~resolution of the S140 (left) and Mon
R2 (right) massive star forming regions. Known sources and features have been
annotated (see text). A vertical band of emission in the S140 image at abscissa $\sim$0\arcsec~is an image artefact.}
\label{morph}
\end{figure}

\subsection{The mid-IR morphology of the S140 region} 
At a distance 910\,pc, the S140 massive star forming region is
arguably the best example of the initial stages of massive star
formation in the solar neighbourhood \cite{1974QB4.V5v14n12...}. It
consists of a cluster of at least three luminous near-IR sources
without optical counterparts \cite{1979ApJ...232L..47B}, and strong outflow activity. Detailed studies have demonstrated that the
source IRS\,1 consists of a disk found at a position angle of $\sim44\degr$,
perpendicular to the CO outflow and monopolar reflection nebula
\cite{2006ApJ...649..856H,2008ApJ...673L.175J}.

The left panel of Fig.\,\ref{morph} reveals a wealth of features associated
with objects previously identified at different wavelengths. Discrete peaks in
mid-IR emission are found at the positions of IRS\,1, 2, and 3. IRS\,1 is
symmetric, whereas IRS\,2 and IRS\,3 are elongated. The latter
is found to be a triple system \cite{2001A&A...378..539P} and our image partly
resolves the secondary object (IRS3b) at a distance of $0.75\arcsec$~east of
the primary source (IRS3a).

The COMICS image also shows patches of diffuse emission that are found
coincident with the radio sources VLA4 and NW \cite{1989ApJ...346..212E}. This
is the first time that the two radio sources are seen in the
mid-IR. Previously \cite{1989ApJ...346..212E}, they were found coincident with
the brightest parts of very extended near-IR nebulosity. 
Noticeable is the spectacular arc of mid-IR emission labelled `I'. This dust
emission structure concurs with an emission arc at $K$-band. The causes for
the formation of this relatively large structures is unclear, as they do not
align with the principal IRS1 outflow direction.
Finally, the curved wisp of mid-IR emission at
$\sim3\arcsec$ from IRS3 corresponds to feature ``F'' of
\cite{2001A&A...378..539P}. This feature demonstrates strongly polarised
emission and strong $\rm H_{2}$ line emission. The structure is possibly
created by outflow activity from IRS3. Perhaps surprisingly, we find no warm dust emission associated
with the two submm emission peaks \cite{1995A&A...298..894M} that
lie within our COMICS field. The S140 region shows 24.5\,\micron~emission to have a
diffuse character when it apparently traces shocked dense material, but also
when it is associated with radio sources.

\subsection{The mid-IR morphology of the Mon R2 region}
The Mon R2 region displays massive stars in apparently three successive stages
of formation. The earliest stage is identified with the complex mm emission that
suggests the formation of cold cores away from the IR sources
\cite{1997ApJ...487..346G}. The discrete, and more evolved, MYSOs in the
COMICS image are separated by a ridge of mid-IR emission.  At the maximum
intensity along the mid-IR ridge, just on the inside close to the smaller of
the two crosses, the diffuse nebulosity corresponds to the location of an
extended HII region, the most advanced stage in this region, harbouring the
near-IR source IRS1. Again the HII region does not have a discrete
24.5\,\micron~counterpart. 

The large-scale mid-IR shell structure has a near-IR counterpart \cite{1994ApJ...425..707H}. The large cross in Fig.\,\ref{morph}
corresponds to the centre of the CO J=3-2 ``hole'' as identified in
\cite{1997ApJ...487..346G}. Its morphology corresponds roughly to the extend
of the HII region \cite{1985A&A...152..387M}.  If we compare the
mid-IR image with these maps, we can identify the 24.5\,\micron~emission with
the walls of the ionised region.


\section{MYSO envelope structure from the SED and the 24.5$\mu$m intensity profile}
The lack of asymmetries seen in the discrete and partially resolved MYSO
envelopes warrants the use of simple spherical radiative transfer models.  The
envelopes can be described by powerlaws in density
$\rho=\rho_{\circ}\,r^{-n}$, where the power index $n$ reveals the physics
governing the structure. In this section we discuss in some detail two
examples of simultaneously modelling the 24.5\,\micron~emission and SED with
aiming to extract the density power index from the observables. Simultaneous
modelling seeks consistency between the spatial and spectral information. The
size scale of the envelope depends strongly on \lbol~determined from the
far-IR and submm flux levels. The wavelength region $<10$\,\micron~is ignored as
various components other than the envelope may contribute.  Two very different
examples are presented: S140\,IRS1 and M8E-IR. The case of the dominant MYSO
in the Mon\,R2 region (IRS3) is nearly identical to S140\,IRS1, and its
modelling alongside other details of the modelling of our complete MYSO sample
can be found in de Wit et al. 2008 (subm).

\begin{figure} 
  \center{
    \includegraphics[width=5.5cm,height=7.3cm,angle=90]{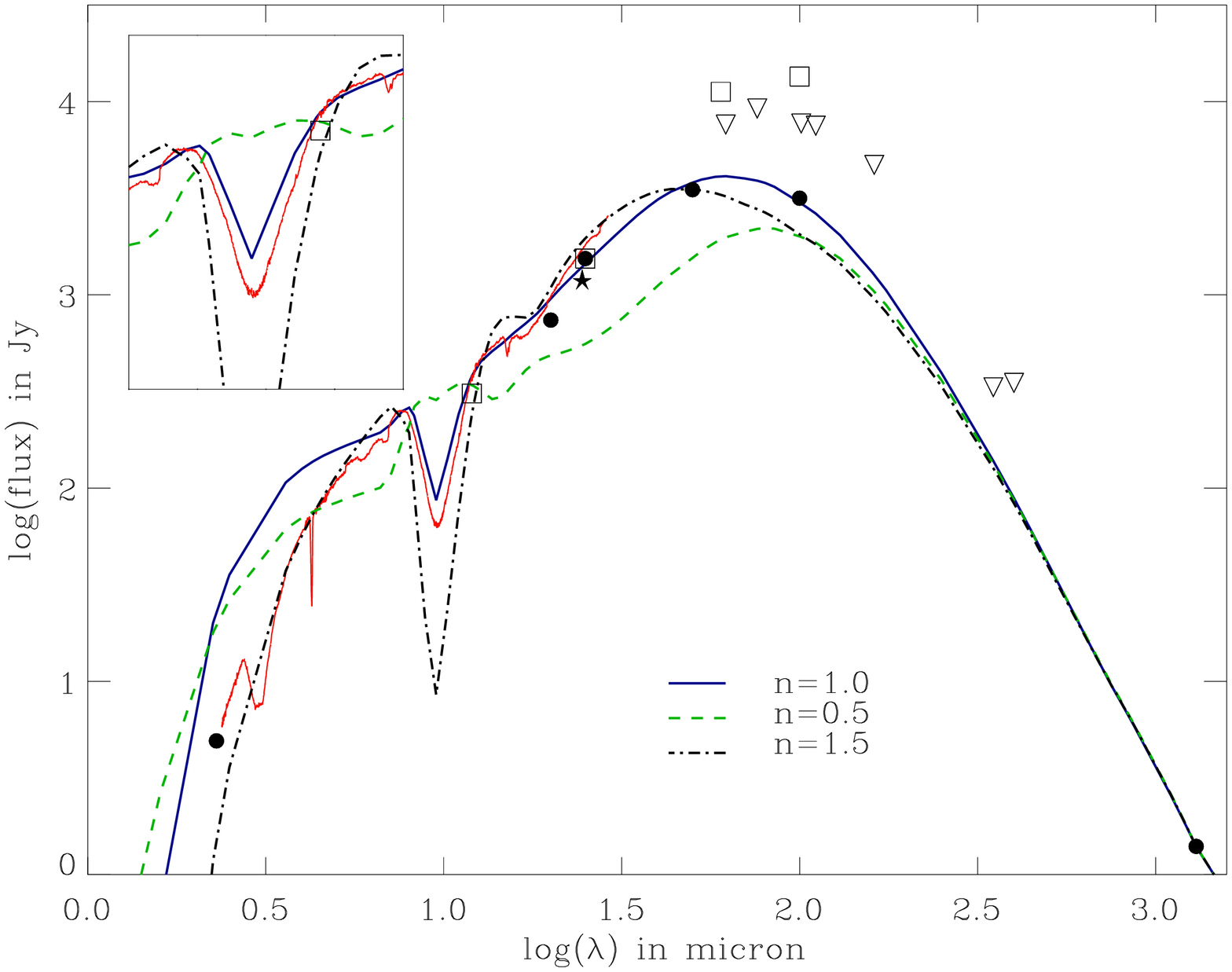} 
    \includegraphics[width=5.5cm,height=7.3cm,angle=90]{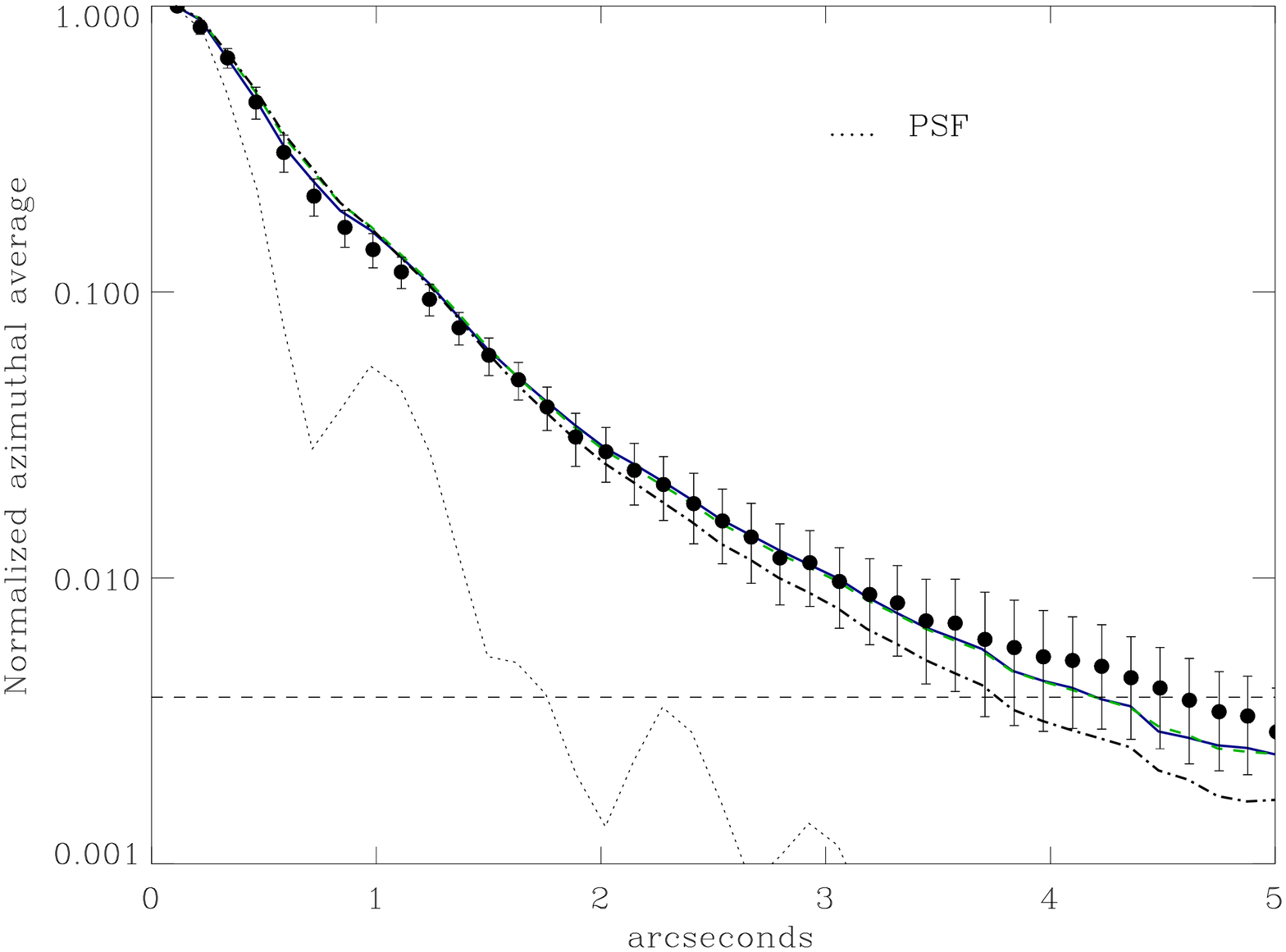} 
  }
  \caption[]{Observations and modelling of S140\,IRS1. {\it Left:} Asterisk: COMICS flux, filled circles: KAO 50 and
    100\,\micron~\cite{1986ApJ...309...80L}, 1.3mm SEST
    \cite{1991A&A...252..801G}. Open symbols (not used in model fit): IRAS and submm
    data from \cite{1991A&A...252..801G,1983ApJ...271..625S}. 
{\it Right:} 24.5\,\micron~intensity profile. The best simultaneous fitting model
has a density profile with a $n=1.0$ powerlaw.}  
\label{s140m}
\end{figure}

\begin{figure} 
    \includegraphics[width=5.5cm,height=7.3cm,angle=90]{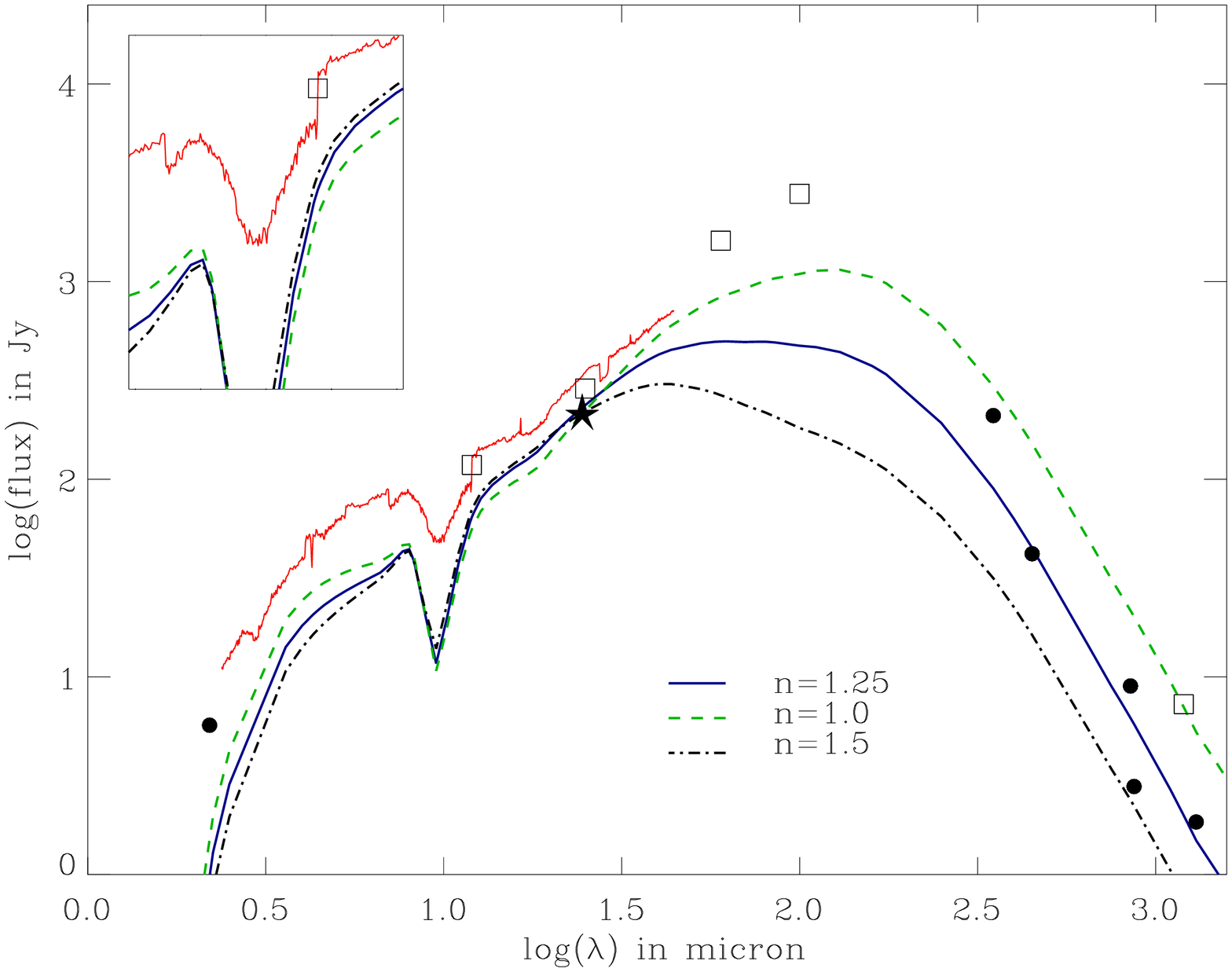} 
    \includegraphics[width=5.5cm,height=7.3cm,angle=90]{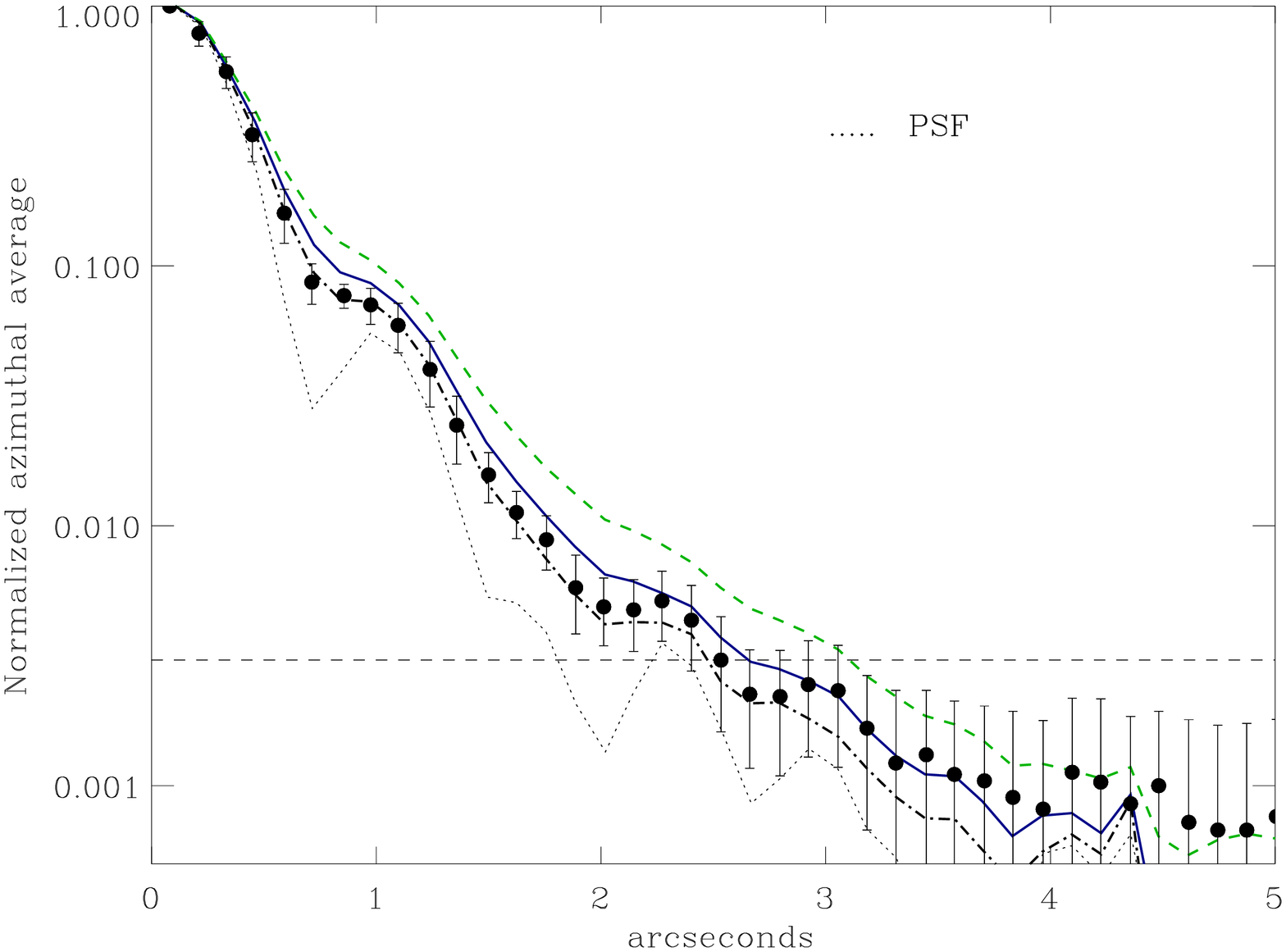} 
    
    \caption[]{Observations and modelling of M8E-IR {\it Left:} Asterisk: COMICS flux, other data from  
\cite{2002ApJS..143..469M,1991A&A...252..801G}.
      The level of (sub)mm continuum flux is uncertain, due to beam size effects. FWHM varies between 9\arcsec (870\,\micron) 
and 30\arcsec. For reference the 1.2\,mm data point by \cite{2006A&A...447..221B} with a FWHP of 26\arcsec~(open
      square). Short wavelength data by \cite{1985ApJ...298..328S}. {\it
        Right:} The 24.5\,\micron~intensity profile. Best fitting model has a density powerlaw with
      $n=1.25$ index.}  
    \label{m8em}
\end{figure}

S140 IRS1 and M8E-IR are the brightest and dominant mid-IR sources in the
respective regions. The SED continuum measurements from the literature are supplemented with ISO-SWS spectra that
provide strong constraints for the total dust optical depth through the envelope
by means of the 9.7\,\micron~absorption feature. Data and models are presented
in Fig.\,\ref{s140m} and \ref{m8em}, models are fit to SED's filled circles and ISO data.

S140\,IRS1 is extended and clearly resolved. The length of the intensity
profile errorbars cover the range of pixel values at each radial
distance bin. Their small values indicate that IRS1 can be considered to be
symmetric to first order. Various density distributions are able to fit
the intensity profile as well. The simultaneous fit to the SED excludes
however the shallower $n=0.5$ or the steeper $n=1.5$ powerlaws and $n=1.0$ is
preferred. A steeper radial profile requires high optical depths and produces
silicate absorption that is deeper than observed. 

M8E-IR is compact and only marginally resolved at 24.5$\mu$m. The spherical
model that fits the intensity profile best has an $n=1.5$ powerlaw. It
predicts however too little flux at long wavelengths for the envelope outer
radii we initially assumed. In order to increase the (sub)mm flux levels
without increasing the bolometric luminosity we make the envelope larger
by adding cool material at the outer fringes.  Depending on what the true (sub)mm
flux is, $n=1.25$ to $n=1.5$ are preferred with a relatively large outer radius
of 1pc. The flux level of the ISO-SWS spectrum can however not be
attained. The models are forced to fit the COMICS flux data point given 
the lack of unambiguous flux level at these wavelengths.  We
conclude that a spherical model is capable of reproducing both the
24.5\,\micron~profile and SED simultaneously for a dust density radial
distribution of $n=1.25$ and an outer radius of 3000 times the inner
radius. On scales of 1000-10\,000\,AUs, the 350\,\micron~intensity profile 
is best represented by models with $n=1.75$ radial powerlaw density distributions
\cite{2002ApJS..143..469M}. 
\newline

In summary, for 60\% of our MYSO sample spherical models are capable of
satisfactorily reproducing the SED and intensity profile simultaneously. The
fit includes a correct prediction for the depth of the silicate absorption
profile at 9.7$\mu$m, the flux levels at 24.5\,\micron~and at (sub)mm
wavelengths, and of course the 24.5\,\micron~intensity profile. The wavelength
region that is least constrained by the data is the 100\,\micron~wavelength
region (the peak of the SED), where in most cases the IRAS data points
constitute the only flux measurement, however they have too coarse an
angular resolution in our analysis.

Two objects (out of 10) can not be modelled with spherical models: AFGL\,2591 and
S255 IRS\,3. The \lbol~and the size scale of the envelope
are incompatible in these cases. M8E-IR actually poses similar problems regarding
the mid-IR flux shortward of 25$\mu$m. In fact, there is excess mid-IR
flux with respect to the spherical model that can only be accounted for if a
higher \lbol~is adopted, a solution denied by the intensity profile. It would
suggest that mid-IR emission from these sources is not fully dominated by the
envelope.

Noteworthy is that the MYSOs that have the poorest fits to spherical models
(40\%), are the ones represented by relatively steep radial density distributions.
It is probably not a coincidence that all these sources have evidence for cavity
wall emission or outflow activity along the line of sight. Their poor model fits
probably reflect the inadequacy of spherical models.

The remaining six objects are well reproduced by spherical models and four of them have a
preferred $n=1.0$ powerlaw. Two cases show even shallower density
profiles. However in case of IRAS\,20126 emission is most likely dominated by disk rather and/or outflow cavity emission \cite{2007ApJ...654L.147D}. Only little background information is 
available for the second flat source AFGL\,437S. We can
therefore conclude tentatively that when relatively steep spherical models fit
MYSOs SEDs and intensity profiles, the source SED is probably not uniquely
determined by envelope emission but partially by inhomogeneities in the envelope
(cavity walls). Otherwise, for those
MYSO where spherical models are capable of reproducing simultaneously
the 24.5\,\micron~intensity profile and the SED, radial density profiles power
index of $n=1.0$ on scales of 1000\,AU are preferred.

\section{Envelope emission at milli arcsecond scale resolution with VLTI/MIDI}
Resolving the circumstellar environment of MYSO on scales of 0.05
arcseconds can be attained in interferometric mode with VLT and the
MIDI instrument. MIDI operates at 10\,\micron~and delivers spectrally
dispersed visibilities. Various components in the circumstellar
environment of a MYSO may contribute to emission at wavelengths
$<30$\,\micron.  Near-IR photons may originate either from the stellar
surface, an inner dust truncation structure or from an accretion
disk. They can easily scatter and escape through existing
inhomogeneities in the spherical envelope \cite{2000A&A...353..211H}
and still suffer extinction from any foreground molecular cloud
material, e.g. \cite{2005ApJ...635..452D}.  Under favourable
inclinations, mid-IR radiation from the surface of cavities sculpted
by polar outflows can be seen. In this case, the mid-IR photons are
emitted by warm dust particles that have a clear line-of-sight to the
star, e.g. \cite{2007ApJ...654L.147D}. MIDI has the potential of
revealing the geometry of the various interacting components involved
in shaping a massive star (see also Linz et al. in these proceedings).
We executed a MIDI program on the MYSO W33A aimed at resolving the
10\,\micron~emission, and determining which component of the MYSO environment dominates it.

\begin{figure} 
  \includegraphics[width=5.5cm,height=7.3cm,angle=90]{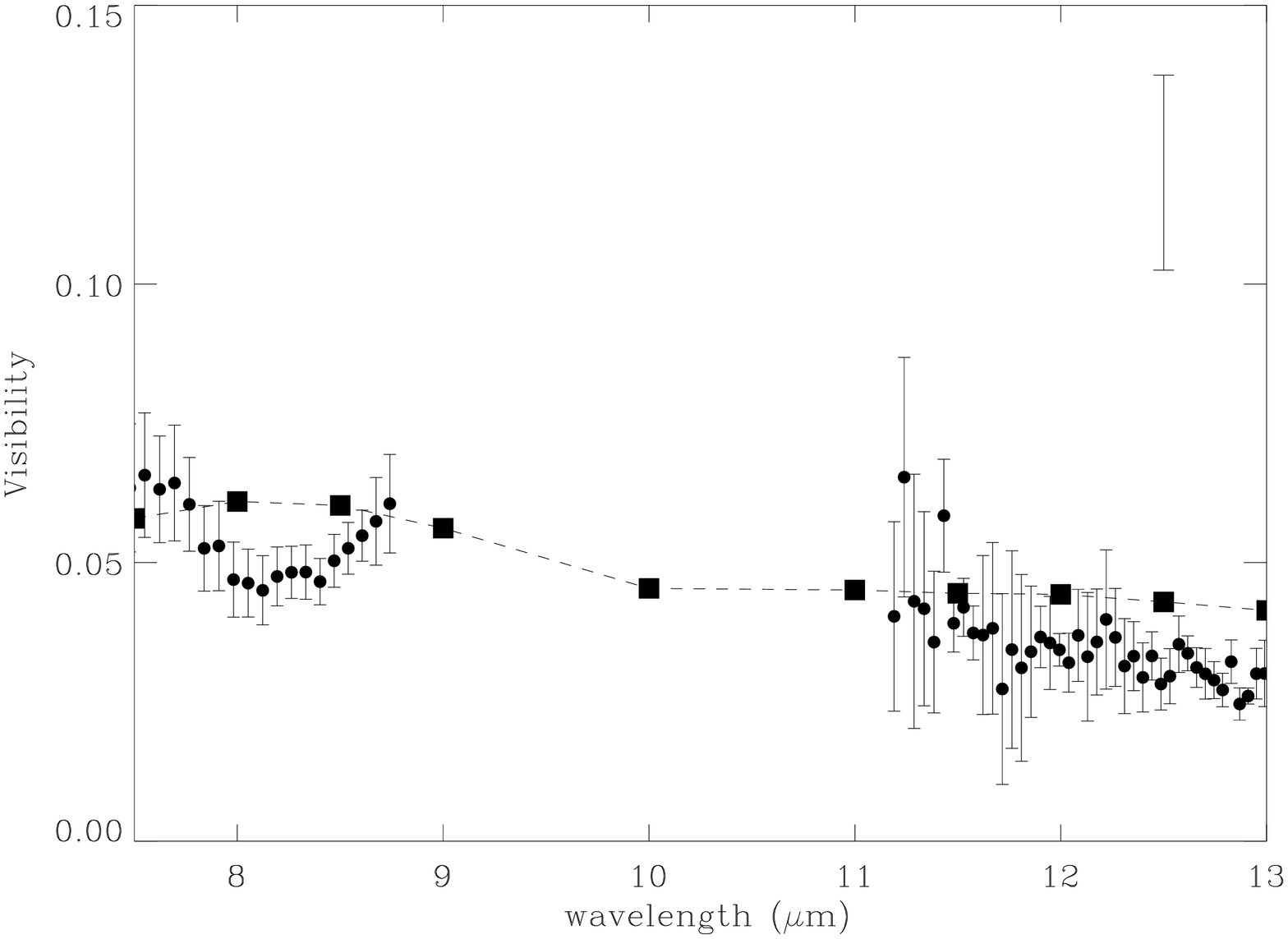} 
  \includegraphics[width=5.5cm,height=7.3cm,angle=90]{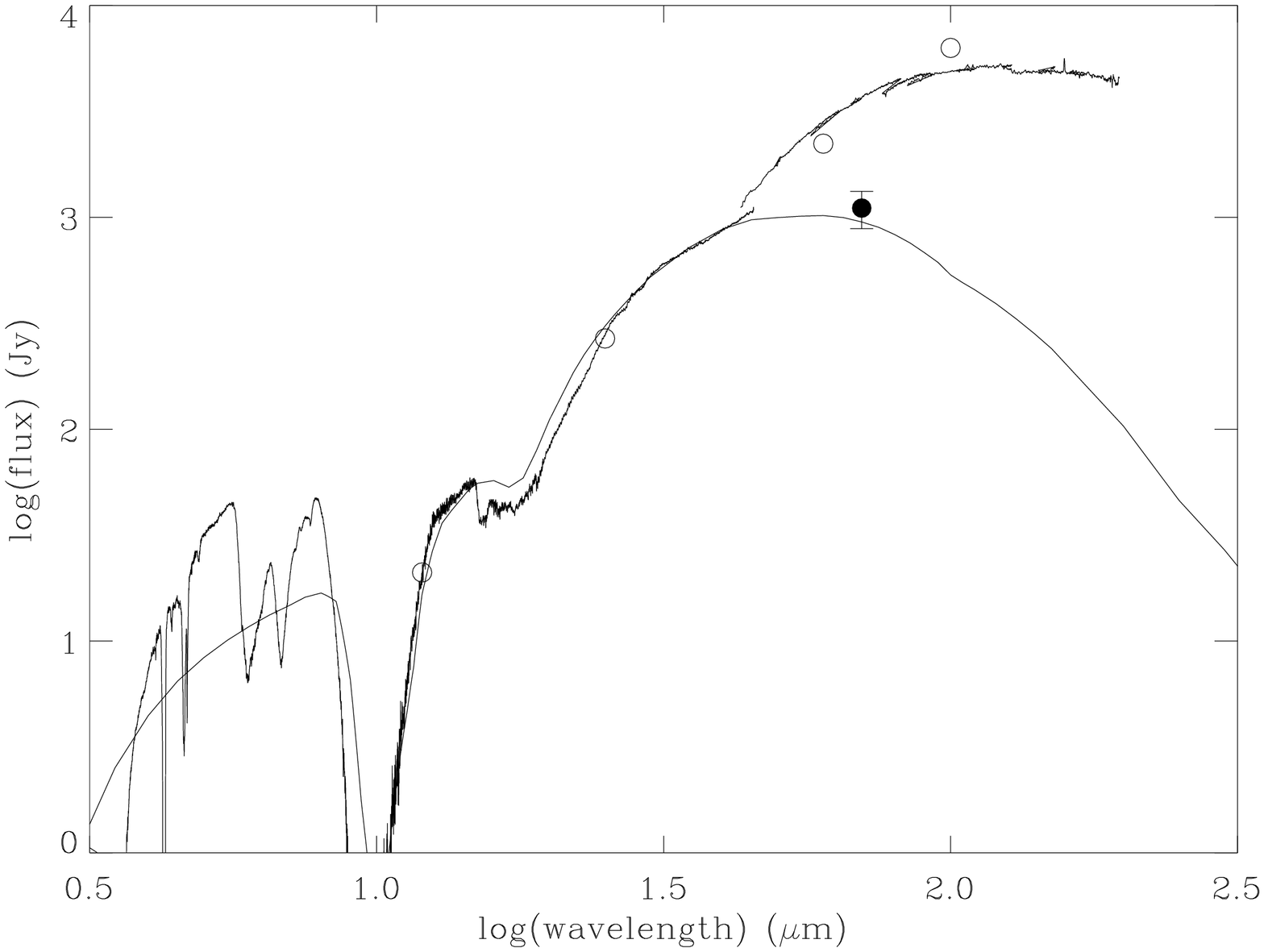} 
  \caption[]{{\it Left:} MIDI visibilities of W33A (with errorbars). The dashed
    line is best fitting DUSTY model. {\it Right:} Simultaneous model fit to
    the SED. Shown are ISO spectra, and IRAS data. The data point with errorbar is from Spitzer MIPS.}  
  \label{midi}
\end{figure}

The targeted MYSO, W33A, has a kinematic distance of 3.8 kpc and a luminosity as derived
from IRAS fluxes of $\rm 1 \times 10^{5} L_{\odot}$ \cite{2004A&A...426...97F}. It has weak, compact, optically
thick radio continuum emission \cite{1996ApJ...465..363R,2005A&A...437..947V} and broad ($\sim$100 km s$^{-1}$),
single-peaked H I recombination emission lines \cite{1995MNRAS.272..346B} consistent with an ionised
stellar wind origin. IR images from 2MASS and Spitzer's GLIMPSE survey
clearly show a large scale monopolar nebula emerging to the SE.

Fig.\,\ref{midi} presents the measured MIDI visibilities (with errorbars) and
the SED. W33A shows a particularly deep silicate absorption feature. At the central
wavelength of the feature no flux was recorded, and the actual depth is
unknown. This is the reason why the visibility spectrum in Fig.\,\ref{midi}
does not show any measurement between 9 and 11\,\micron. 
The visibilities have a declining trend with wavelength. If we would
represent the emission by a Gaussian emitting distribution then the FWHM size
increases from 30\,mas (115\,AU) at 8$\mu$m to 60\,mas (230\,AU) at 13$\mu$m.

Modelling the 10\,\micron~emission is again performed with 1D DUSTY. Our basic
aim is to probe whether the spatial and spectral information can be accounted
for by envelope emission alone. Fitting 1D DUSTY models to the visibilities
and SED of W33A shows that for nominal stellar parameters, the size scales of the
emission region are too large and produce the wrong trend with
wavelength. Relatively shallow radial density distributions with power indices
between $n=0.5$ and $n=1.0$ reproduce better the observed dependency of size with
wavelength. The deep silicate feature observed in the flux spectrum is better
reproduced by dust models with warm Ossenkopf silicates that have an increased
silicate over graphite ratio. For all models
however the bolometric luminosity and $T_{\rm eff}$ of the central object need
to be reduced. This reduction in \lbol~is in fact in better agreement with
70\,\micron~Spitzer photometry, as can be seen in Fig.\,\ref{midi}.
In summary, the reasonable fit with spherical models thus suggest that in case of W33A
dominant component at 10\,\micron~emission is the envelope.

The MIDI observations thus reveal that any circumstellar accretion disk
structure should probably be smaller than the inner extent of the envelope,
i.e. $\sim$100\,AU.  In addition, the reduction in $T_{\rm eff}$ would imply
that the central star may be swollen.  These results are consistent with the
MIDI findings for M8E-IR as presented by Linz et al. (these proceedings). A
bloated star would support the idea of induced swelling by high-mass accretion rates
causing very large radii of the central object \cite{2007arXiv0711.4912H,2008ASPC..387..255H}.

\section{Concluding remarks}
High angular resolution mid-IR observations are starting to give a clear
and detailed view of how the circumstellar material of a MYSO is organised. 
In this contribution we have presented an analysis of circumstellar emission of
MYSOs in the mid-IR at two different wavelengths (24.5\,\micron~and 10\,\micron)
and two different resolutions (0.6\arcsec~and 0.05\arcsec). We have found that
when the emission at these wavelengths is dominated by dust emission from the
envelope the radial density profiles can be described by a powerlaw
of the form $\rho=\rho_{0}\,r^{-n}$ with a power index of approximately
$n=1.0$. Steep powerlaws $n\geq1.5$ are excluded as they require high optical
depth denied by the silicate absorption profile. Steep powerlaws are found but
only in those cases where their is evidence for a contribution by the outflow
cavity to the mid-IR emission. We note that our results for the radial density
distribution does not negate the findings of previous studies conducted in the
(sub)mm wavelength region. For example \cite{2002ApJS..143..469M} claim a
value of of $1.8\pm0.4$ from analysis for 350\,\micron~intensity
profiles. However our results seem to be more in line with lower values found
by \cite{2000ApJ...537..283V}. The powerlaw dependence of density with radius
is predicted to have a powerindex of $n=1.0$ when the envelope is supported by 
turbulence \cite{1989ApJ...342..834L}, although on the scales probed with MIDI
rotational flattening may start to play a significant role.

\section*{References}
\providecommand{\newblock}{}

\end{document}